\documentstyle[preprint,aps,eqsecnum]{revtex}
\begin{document}
\preprint{UAHEP 958}
\draft
\title{New Perturbative Solutions of the Kerr--Newman Dilatonic
Black Hole Field Equations}
\author{R. Casadio\cite{cas}, B. Harms and Y. Leblanc}
\address{Department of Physics and Astronomy,
The University of Alabama\\
Box 870324, Tuscaloosa, AL 35487-0324}
\author{P.H. Cox}
\address{Physics Department, Texas A\&M University-Kingsville\\
Kingsville, TX 78363}
\maketitle
\begin{abstract}
This work describes new perturbative solutions to the classical,
four-dimensional Kerr--Newman dilaton black hole field equations.
Our solutions do not require the black hole to be slowly rotating.
The unperturbed solution is taken to be the ordinary Kerr solution,
and the perturbation parameter is effectively the square of the
charge-to-mass ratio $(Q/M)^2$ of the Kerr--Newman black hole.
We have uncovered a new, exact conjugation (mirror) symmetry for
the theory, which maps the small coupling sector to the
strong coupling sector ($\phi \to -\phi$).
We also calculate the gyromagnetic ratio of the black hole.
\end{abstract}
\section{Introduction}
In recent years a great deal of effort has gone into the study
of black hole solutions in both the classical and the
quantum theory of gravity.
In particular the role of black holes in the quantized theory
of gravity has been extensively investigated, but no consensus
has been reached on what this role is.
\par
In several recent articles \cite{hl1,hl2,chl1,hl3,hl4,hl5,hl6,hl7},
we have analyzed this problem, especially in the context of the
Schwarzschild black hole solution, and have come to the conclusion
that quantum black holes can be identified as particle excitations of
quantum extended objects such as p-branes or strings \cite{hl1}.
From this point of view quantum black holes should be treated as
particles possessing mass, charge and spin.
In addition to these properties, theories of extended objects have
gravitational sectors with both tensor and scalar fields.
So a consistent treatment of the gravitational interactions of such
objects includes the latter, the dilaton field, as well as the
former.
\par
Our objective is to find the general classical solutions to the field
equations in four dimensions for a massive, charged, and spinning
black hole interacting with a dilaton.
The proper action for such a system is the Einstein--Maxwell action
enriched with a dilaton field.
The solutions of the field equations are parametrized by the
mass $M$, the charge $Q$, the (spin) angular momentum $J$ and the
dilaton parameter $a$.
\par
At present there is no known solution for an arbitrary value of $a$.
Exact solutions have been found, however, for the cases $a = 0 $
and $a = \sqrt{3}$.
While the former is just the classic Kerr--Newman solution\cite{chan},
the latter is called the Kaluza--Klein solution since it can be shown to
be equivalent to a compactification of the 5-dimensional vacuum Einstein
equations \cite{horne,gibb}.
Another exact solution has been found by Sen\cite{sen}, but this
solution requires the incorporation of an additional axion
field into the action.
Perturbative solutions are known in the case of arbitrary
$a$\cite{horne,camp}.
The perturbation parameter is the black hole angular momentum $J$
per unit mass, and the corresponding solutions are for slowly
rotating black holes.
\par
Our perturbative expansion, on the other hand, is performed
in terms of the charge-to-mass ratio of the black hole and therefore
is not restricted to the slowly rotating approximation.
The unperturbed solution in our method is the neutral
four-dimensional Kerr space--time\cite{chan}.
\par
In Section II below we present a derivation of the relevant
forms for the field equations of the dilaton Kerr--Newman
problem.
Our notation follows that of Chandrasekhar \cite{chan} very
closely, except that our  convention is
$eta_{ij} = {\rm diag}(-1, 1, 1, 1)$.
In Section III we discuss an exact conjugation (mirror) symmetry
of the field equations, which maps small to strong gravitational
coupling, $\phi \to -\phi$\cite{schw}.
In Section IV we do perturbation theory about small electric charge
$Q^2$ and obtain explicit solutions to order $O(Q^4)$.
In Section V we compare our results to those obtained in the
slowly rotating approximation, including a calculation of the
black hole gyromagnetic ratio\cite{horne,strau}.
\section{Derivation of the field equations}
The action describing the Einstein--Maxwell theory interacting
with a dilaton in four dimensions is ($G = 1$),
\begin{eqnarray}
S = {1\over{16\pi}}\int d^4x
\sqrt{-g}\left[ R - {1\over{2}} (\nabla\phi)^2 -
e^{-a\phi}F^2\right] \; ,
\end{eqnarray}
where $R$ is the scalar curvature, $\phi$ is the dilaton field
and $F_{ij}$ is the Maxwell field.
In tensor components, the corresponding field equations are
\begin{eqnarray}
\left\{\begin{array}{lr}
\nabla^i(e^{-a\phi}F_{ij}) = 0&  \\
& \ \ {\rm (Maxwell)} \\
\nabla_{[ k}F_{ij]} = 0 \; ,& \\
\end{array}\right. \\
& \nonumber \\
\nabla^2\phi = -a e^{-a\phi} F^2 \ \ \ \ {\rm (dilaton)}\; ,&
\end{eqnarray}
and
\begin{eqnarray}
R_{ij} = {1\over{2}} \nabla_i\phi \nabla_j\phi + 2T^
{EM}_{ij} \ \
{\rm (Einstein)}
\; ,
\end{eqnarray}
where $R_{ij}$ is the Ricci tensor, and
the electromagnetic energy--momentum tensor is given as
\begin{eqnarray}
T^{EM}_{ij} = e^{-a\phi} \left[ F_{ik} F^k_j - {1\over{4}} g_{ij}
F^2\right]
\; .
\end{eqnarray}
\par
The general form for the metric which describes a stationary
axisymmmetric space--time is \cite{chan}
\begin{eqnarray}
ds^2 = -e^{2\nu}(dt)^2 + e^{2\psi}(d\varphi - \omega dt)^2 +
e^{2\mu_2}(dr)^2 + e^{2\mu_3}(d\theta)^2
\; ,
\end{eqnarray}
in which $\nu, \psi, \omega, \mu_2, {\rm and}\ \mu_3$ are
functions of $r$ and $\theta$ only.
Following Chandrasekhar\cite{chan}, we shall work in the special frame
of an inertial local coordinate system (tetrad frame) with the basis
1-forms
\begin{eqnarray}
\Omega^0 = e^{\nu} dt \; ;\ \
\Omega^1 = e^{\psi}(d\varphi-\omega dt) \nonumber \\
\Omega^2 = e^{\mu_2}dr\; ; \ \
\Omega^3 = e^{\mu_3}d\theta
\; .
\end{eqnarray}
The coordinates are
\begin{eqnarray}
x^0 = t\; ;\ \ x^1 = \varphi\; ;\ \ x^2 = r\; ; \ \ x^3 = \theta
\; .
\end{eqnarray}
The main advantage of a tetrad frame is that it is locally flat:
\begin{eqnarray}
e^j_a e_{jb} = \eta_{ab} = {\rm diag}(-1,1,1,1)
\end{eqnarray}
The vierbein field corresponding to the frame of Eq.(2.7) is
\begin{eqnarray}
e^i_0 &=& (e^{-\nu}, \omega e^{-\nu},0,0) \nonumber\\
e^i_1 &=& (0,e^{-\psi},0,0) \nonumber \\
e^i_2 &=& (0,0,e^{-\mu_2},0) \nonumber \\
e^i_3 &=& (0,0,0,e^{-\mu_3})
\end{eqnarray}
and
\begin{eqnarray}
e_{0i} &=& (-e^{\nu},0,0,0)\nonumber \\
e_{1i} &=& (-\omega e^{\psi},e^{\psi},0,0)\nonumber \\
e_{2i} &=& (0,0,e^{\mu_2},0)\nonumber \\
e_{3i} &=& (0,0,0,e^{\mu_3})\; .
\end{eqnarray}
Note that latin indices $i, j, k$ are tensor indices,
while $a, b, c$ are tetrad indices.
\par
In the calculations which follow we shall find it convenient to define
\begin{eqnarray}
f_{ab} \equiv F_{ab} e^{-a\phi}
\; ,
\end{eqnarray}
where $F_{ab}$ and $\phi$ are by axisymmetry
functions of $r$ and $\theta$ only.
Maxwell's equations (2.2) in the tetrad basis take the following form
when we group the equations which contain
only total derivatives
\begin{mathletters}
\label{allequations}
\begin{eqnarray}
({e}^{\psi+\mu_2} F_{12}),_3 - (e^{\psi + \mu_3} F_{13}),_2 = 0
\nonumber \\
(e^{\psi + \mu_3} f_{02}),_2 + (e^{\psi + \mu_2}
f_{03}),_3 = 0 \label{equationa}
\end{eqnarray}
and
\begin{eqnarray}
(e^{\nu+\mu_3} f_{12}),_2 + (e^{\nu + \mu_2}
f_{13}),_3 = e^{\psi + \mu_3} f_{02} \omega,_2 +
e^{\psi + \mu_2} f_{03} \omega,_3
\nonumber  \\
(e^{\nu + \mu_2} F_{02}),_3 - (e^{\nu + \mu_3}
F_{03}),_2 = e^{\psi + \mu_2} F_{12} \omega,_3
-e^{\psi + \mu_3} F_{13} \omega,_2 \; .\label{equationb}
\end{eqnarray}
\end{mathletters}
All other components vanish.
\par
Equations (2.13a) are just integrability conditions which
allow us to define ``potentials'' $A$ and $B$ as
\begin{eqnarray}
e^{\psi + \mu_2} F_{12} = A,_2 \; ; \; e^{\psi + \mu_3}
F_{13} = A,_3  \nonumber \\
e^{\psi + \mu_2} f_{03} = - B,_2 \; ; \; e^{\psi + \mu_3}
f_{02} = B,_3
\; .
\end{eqnarray}
\par
Inserting the expressions in Eq.(2.14) into Eqs.(2.13b)
yields the following form for Maxwell's equations,
\begin{eqnarray}
(e^{-\psi+\nu-\mu_2+\mu_3-a\phi}A,_2),_2 + (e^{-\psi+\nu+
\mu_2-\mu_3-a\phi}A,_3),_3 = \omega,_2B,_3 - \omega,_3B,_2
\nonumber \\
(e^{-\psi+\nu-\mu_2+\mu_3+a\phi}B,_2),_2 + (e^{-\psi+\nu+
\mu_2-\mu_3+a\phi}B,_3),_3 = \omega,_3 A,_2 - \omega,_2 A,_3
\; .
\end{eqnarray}
\par
Again in the tetrad frame, the expression for the dilaton
field (Eq.(2.3)) becomes
\begin{eqnarray}
&&e^{-\mu_2}(e^{-\mu_2}\phi,_2),_2 +
e^{-\mu_3}(e^{-\mu_3}\phi,_3),_3
+ e^{-2\mu_2}\phi,_2 (\beta,_2+\mu_3,_2) +
e^{-2\mu_3}\phi,_3(\beta,_3+\mu_2,_3)\nonumber \\
&&= -2a e^{a\phi-2\psi}[e^{-2\mu_2}
({A,_2}^2-{B,_2}^2) + e^{-2\mu_3}({A,_3}^2-{B,_3}^2)]
\; ,
\end{eqnarray}
in which $\beta$ is defined as
\begin{eqnarray}
\beta \equiv \psi + \nu
\; .
\end{eqnarray}
\par
The Einstein equations (2.4) and (2.5) are
\begin{eqnarray}
R_{ab} = {1\over{2}}e^{-\mu_a-\mu_b} \phi,_a\phi,_b + 2T_{ab}
\end{eqnarray}
where $\phi,_0 = \phi,_1 = 0$ and
\begin{eqnarray}
T_{ab} = \left[\eta^{cd}f_{ac}f_{bd} - {1\over{4}}\eta_{ab}
f^2\right]e^{a\phi}
\end{eqnarray}
with
\begin{eqnarray}
f^2 = f_{ab}f^{ab} = 2[-f^2_{02}-f^2_{03}+f^2_{12}+f^2_{13}]
\; .
\end{eqnarray}
\par
In the tetrad basis the non-vanishing Ricci components are\cite{chan}
\begin{eqnarray}
R_{00} &=&
e^{-2\mu_2} [\nu,_2,_2 + \nu,_2(\psi + \nu - \mu_2 + \mu_3),_2]
+e^{-2\mu_3}[\nu,_3,_3 + \nu,_3(\psi + \nu + \mu_2 - \mu_3),_3]
\nonumber \\
&& -{1\over{2}}e^{2(\psi-\nu)}[e^{-2\mu_2} {\omega,_2}^2 +
e^{-2\mu_3} {\omega,_3}^2]
\; ,
\end{eqnarray}
\begin{eqnarray}
R_{11} &= & -e^{-2\mu_2}[\psi,_2,_2 + \psi,_2(\psi + \nu +
\mu_3 - \mu_2),_2]
-e^{-2\mu_3}[\psi,_3,_3 + \psi,_3(\psi + \nu + \mu_2 - \mu_3),_3]
\nonumber \\
&& -{1\over{2}}e^{2(\psi-\nu}[e^{-2\mu_3} {\omega,_3}^2 +
e^{-2\mu_2} {\omega,_2}^2] \; ,
\end{eqnarray}
\begin{eqnarray}
R_{22} &= & -e^{-2\mu_2}[(\psi+\nu+\mu_3),_2,_2 +
\psi,_2(\psi-\mu_2),_2 + \mu_3,_2(\mu_3-\mu_2),_2
+\nu,_2(\nu-\mu_2),_2] \nonumber \\
&&-e^{-2\mu_3}[\mu_2,_3,_3 + \mu_2,_3(\psi+\nu+\mu_2
-\mu_3),_3]
+{1\over{2}} e^{2(\psi-\nu-\mu_2)} {\omega,_2}^2
\; ,
\end{eqnarray}
\begin{eqnarray}
R_{33}& = & -e^{-2\mu_3}[(\psi+\nu+\mu_2),_3,_3 +
\psi,_3(\psi-\mu_3),_3 + \mu_2,_3(\mu_2-\mu_3),_3
+ \nu,_3(\nu-\mu_3),_3]
\nonumber \\
&& -e^{-2\mu_2}[\mu_3,_2,_2+\mu_3,_2(\psi+\nu+\mu_3
-\mu_2),_2]
+ {1\over{2}} e^{2(\psi-\nu-\mu_3)} {\omega,_3}^2 \; ,
\end{eqnarray}
\begin{eqnarray}
R_{01} = {1\over{2}}e^{-2\psi-\mu_2-\mu_3}
[(e^{3\psi-\nu-\mu_2+\mu_3}\omega,_2),_2 +
(e^{3\psi-\nu-\mu_3+\mu_2}\omega,_3),_3]
\; ,
\end{eqnarray}
and
\begin{eqnarray}
R_{23} = && -e^{-\mu_2-\mu_3}[(\psi+\nu),_2,_3 -
(\psi+\nu),_2\mu_2,_3 - (\psi+\nu),_3\mu_3,_2  + \psi,_2\psi,_3 +
\nu,_2\nu,_3] \nonumber \\
&& + {1\over{2}}e^{2\psi-2\nu-\mu_2-\mu_3} \omega,_2\omega,_3
\; .
\end{eqnarray}
Multiplying the $R_{00}$ and $R_{11}$ components by $e^{
\beta+\mu_2+\mu_3}$ and making use of Eq.(2.18), we arrive
at the relations
\begin{eqnarray}
[e^{\mu_3-\mu_2}(e^{\beta}),_2],_2 + [e^{\mu_2-\mu_3}
(e^{\beta}),_3],_3 = 0
\; ,
\end{eqnarray}
and
\begin{eqnarray}
[e^{\beta+\mu_3-\mu_2}(\psi-\nu),_2],_2 +
[e^{\beta+\mu_2-\mu_3}(\psi-\nu),_3],_3
& =& -e^{3\psi-\nu}[e^{\mu_3-\mu_2} {\omega,_2}^2 +
e^{\mu_2-\mu_3} {\omega,_3}^2] \nonumber \\
&& -2e^{-\psi+\nu}[e^{\mu_3-\mu_2}(e^{-a\phi} {A,_2}^2
+e^{a\phi} {B,_2}^2) \nonumber \\
&& + e^{\mu_2-\mu_3}(e^{-a\phi} {A,_3}^2+e^{a\phi} {B,_3}^2)]
\; .
\end{eqnarray}
Eq.(2.27) is essentially the sum $R_{00}+R_{11}$ while
Eq.(2.28) is the difference $R_{00}-R_{11}$.
\par
Corresponding relations can be found for the sum and
difference of $R_{22}$ and $R_{33}$.
Making use of the Einstein relation (Eq.(2.18))
and Eqs.(2.23) - (2.24), the difference is
\begin{eqnarray}
4e^{\mu_3-\mu_2}(\beta,_2\mu_3,_2+\psi,_2\nu,_2)
&-&4e^{\mu_2-\mu_3}(\beta,_3\mu_2,_3+\psi,_3\nu,_3) \nonumber \\
&=& 2e^{-\beta}\left\{[e^{\mu_3-\mu_2}(e^{\beta}),_2],_2
-[e^{\mu_2-\mu_3}(e^{\beta}),_3],_3\right\} \nonumber \\
&& -e^{2(\psi-\nu)}[e^{\mu_3-\mu_2} {\omega,_2}^2 -
e^{\mu_2-\mu_3} {\omega,_3}^2] \nonumber \\
&& + (e^{-2\mu_2} {\phi,_2}^2 - e^{-2\mu_3} {\phi,_3}^2)
\,e^{\mu_2+\mu_3} \nonumber \\
&& +4e^{-2\psi}[e^{\mu_3-\mu_2}(e^{-a\phi} {A,_2}^2
+ e^{a\phi} {B,_2}^2) \nonumber \\
&& -e^{\mu_2-\mu_3}(e^{-a\phi} {A,_3}^2 + e^{a\phi} {B,_3}^2)]
\; ,
\end{eqnarray}
while the sum is
\begin{eqnarray}
R_{22}+R_{33} = {1\over{2}}(e^{-2\mu_2} {\phi,_2}^2 +
e^{-2\mu_3} {\phi,_3}^2)
\; .
\end{eqnarray}
\par
Finally, the $R_{01}$ component leads to the relation,
\begin{eqnarray}
(e^{3\psi-\nu-\mu_2+\mu_3}\omega,_2),_2 +
(e^{3\psi-\nu+\mu_2-\mu_3}\omega,_3),_3
= 4(A,_2B,_3 - A,_3B,_2)
\; ,
\end{eqnarray}
while the $R_{23}$ component gives
\begin{eqnarray}
\beta,_2,_3 -&& \beta,_2\mu_2,_3 - \beta,_3\mu_3,_2 +
\psi,_2\psi,_3 + \nu,_2\nu,_3 \nonumber \\
&& = {1\over{2}}e^{2(\psi-\nu)}\omega,_2\omega,_3
-{1\over{2}}\phi,_2\phi,_3
-2e^{-2\psi}(e^{-a\phi}A,_2A,_3 +
e^{a\phi}B,_2B,_3)
\; .
\end{eqnarray}
\par
The final form for the field equations is arrived
at by means of the definitions,
\begin{eqnarray}
& \chi \equiv e^{-\psi+\nu}\; ,\ \ \
\Delta \equiv e^{2(\mu_3-\mu_2)}
\nonumber \\
& \rho^2 \equiv e^{2\mu_3} \Longrightarrow \;
e^{2\mu_2} = \strut\displaystyle{\rho^2\over{\Delta}}&
\nonumber \\
& \Psi \equiv \strut\displaystyle{e^{\beta}\over{\chi}} =
e^{2\psi}  \Longrightarrow e^{2\nu} = \chi e^{\beta}
\; ; \ \  (\beta = \psi + \nu)
\; .&
\end{eqnarray}
For convenience we also define
\begin{eqnarray}
&\mu \equiv \cos\theta &\nonumber \\
& \delta \equiv 1 - \mu^2 = \sin^2\theta
\; . &
\end{eqnarray}
\par
Eq.(2.27) is now readily solved,
\begin{eqnarray}
e^{\beta(r,\theta)} = \sqrt{\delta(\theta)\Delta(r)} =
\sqrt{\Delta(r)}\sin\theta
\end{eqnarray}
where $\Delta(r)$ is quadratic,
\begin{eqnarray}
\Delta(r) & = &r^2-2Mr + M_0^2 \nonumber \\
& = &(r-r_+)(r-r_-)
\end{eqnarray}
with
\begin{eqnarray}
r_{\pm} = M \pm \sqrt{M^2 - {M_0}^2}
\ .
\end{eqnarray}
Such a solution is also found in the case of the Kerr and
Kerr--Newman problems \cite{chan}.
Only the value of $M_0$ differs and
its present value will be determined in the following section.
\par
Recalling the definitions in Eqs.(2.33)--(2.34), we obtain the
relations,
\begin{eqnarray}
e^{2\psi} = \Psi = {\sqrt{\delta\Delta}\over{\chi}}\; ,
\ \ \ \
e^{2\nu} = \chi\sqrt{\delta\Delta}
\; .
\end{eqnarray}
The expressions
\begin{eqnarray}
&\partial_3=\partial_{\theta} = - \sqrt{\delta}\partial_{\mu}
\; ,
\nonumber \\
&\partial_2 = \partial_r \; ,
\end{eqnarray}
allow us to write the field equations as
\begin{eqnarray}
\left[{\Delta\over{\Psi}}e^{-a\phi}A,_r\right],_r +
\left[{\delta\over{\Psi}}e^{-a\phi}A,_{\mu}\right],_{\mu}
=&& -\omega,_rB,_{\mu} + \omega,_{\mu}B,_r \nonumber \\
&& \\
\left[{\Delta\over{\Psi}}e^{a\phi}B,_r\right],_r +
\left[{\delta\over{\Psi}}e^{a\phi}B,_{\mu}\right],_{\mu}
=&& \omega,_rA,_{\mu} - \omega,_{\mu}A,_r \; , \nonumber
\end{eqnarray}
for
{\em Maxwell's equations},
\begin{eqnarray}
(\Delta\phi,_r),_r + (\delta\phi,_{\mu}),_{\mu} = -
{2a\over{\Psi}}&& \left[\Delta(e^{-a\phi} {A,_r}^2 -
e^{a\phi} {B,_r}^2)
+ \delta(e^{-a\phi} {A,_{\mu}}^2 - e^{a\phi} {B,_{\mu}}^2)
\right]
\; ,
\end{eqnarray}
for the {\em dilaton}, and
\begin{eqnarray}
\left[\left({\Delta\Psi,_r\over{\Psi}}\right),_r + \left(
{\delta\Psi,_{\mu}\over{\Psi}}\right),_{\mu}\right]
&=& -{2\over{\Psi}}\left[\Delta(e^{-a\phi} {A,_r}^2 +
e^{a\phi} {B,_r}^2)
+\delta(e^{-a\phi} {A,_{\mu}}^2 + e^{a\phi} {B,_{\mu}}^2)\right]
\nonumber \\
&& - \left[\Delta \left({\Psi\over{\Delta}}\omega,_{\mu}\right)^2 +
\delta\left({\psi\over{\delta}}\omega,_r\right)^2 \right] \; ,
\end{eqnarray}
for the $R_{00}-R_{11}$ component of {\em Einstein's equation} (2.28),
\begin{eqnarray}
\left[\left({\Psi\omega,_r\over{\delta}}\right),_r +
\left({\Psi\omega,_{\mu}\over{\Delta}}\right),_{\mu}
\right] = && {4\over{\Psi}}\left[A,_{\mu}B,_r -
A,_r B,_{\mu}\right] \nonumber \\
&& -\left[\left({\Psi,_r\over{\Psi}}\right)\left(
{\Psi\omega,_r\over{\delta}}\right) + \left({\Psi,_{\mu}
\over{\Psi}}\right)\left({\Psi\omega,_{\mu}\over{\Delta}}
\right)\right] \; ,
\end{eqnarray}
for the $R_{01}$ component (cf. Eq.(2.31)), and finally,
\FL
\begin{eqnarray}
&&-{\Delta\over{2}}\left[
\left({\Psi,_r\over{\Psi}}\right)^2 +
\left({\Psi\omega,_{\mu}\over{\Delta}}\right)^2\right] +
{\delta\over{2}}\left[\left({\Psi,_{\mu}\over{\Psi}} \right)^2
+ \left({\Psi\omega,_r\over{\delta}}\right)^2\right] \nonumber \\
&&+ \Delta,_r\left({\rho,_r\over{\rho}} + {1\over{2}}
{\Psi,_r\over{\Psi}}\right) - \delta,_{\mu}\left(
{\rho,_{\mu}\over{\rho}} + {1\over{2}}{\Psi,_{\mu}\over{\Psi}}\right)
\nonumber \\
&& = 2 + {1\over{2}}\left(\Delta\phi,_r^2 - \delta\phi,_{\mu}^2\right)
+ {2\over{\Psi}}\left[\Delta(e^{-a\phi}A,_r^2 + e^{a\phi}B,_r^2)
-\delta(e^{-a\phi}A,_{\mu}^2 + e^{a\phi}B,_{\mu}^2)\right] \; ,
\end{eqnarray}
\begin{eqnarray}
2\left[\left({\Delta\rho,_r\over{\rho}}\right),_r +
\left({\delta\rho,_{\mu}\over{\rho}}\right),_{\mu}\right]&
= &\Delta,_r\left({\rho,_r\over{\rho}} + {1\over{2}}
{\Psi,_r\over{\Psi}}\right) + \delta,_{\mu}\left(
{\rho,_{\mu}\over{\rho}} + {1\over{2}}{\Psi,_{\mu}
\over{\Psi}}\right) \nonumber \\
&& -{\Delta\over{2}}\left[\left({\Psi,_r\over{\Psi}}\right)^2
- \left({\Psi\omega,_{\mu}\over{\Delta}}\right)^2\right]
- {\delta\over{2}}\left[\left({\Psi,_{\mu}\over{\Psi}}
\right)^2 - \left({\Psi\omega,_r\over{\delta}}\right)^2
\right]
\nonumber \\
&&-{1\over{2}}\left(\Delta {\phi,_r}^2 + \delta {\phi,_{\mu}}^2
\right)
\; ,
\end{eqnarray}
\begin{eqnarray}
\left({\Psi,_r\over{\Psi}}\right)\left({\Psi,_{\mu}
\over{\Psi}}\right) - \left({\Psi\omega,_r\over{\delta}}\right)\left(
{\Psi\omega,_{\mu}\over{\Delta}}\right) &+&
{1\over{2}}{\Delta,_r\over{\Delta}}{\delta,_{\mu}
\over{\delta}}-
{\Delta,_r\over{\Delta}}\left({\rho,_{\mu}\over
{\rho}} + {1\over{2}}{\Psi,_{\mu}\over{\Psi}}\right)
- {\delta,_{\mu}\over{\delta}}\left({\rho,_r
\over{\rho}} + {1\over{2}}{\Psi,_r\over{\Psi}}\right)
\nonumber \\
&& = -{4\over{\Psi}}\left[e^{-a\phi}A,_rA,_{\mu} +
e^{a\phi}B,_rB,_{\mu}\right] -\phi,_r\phi,_{\mu} \; ,
\end{eqnarray}
corresponding to the difference $R_{22} - R_{33}$ (Eq.(2.29)), the sum
$R_{22}+R_{33}$ (Eq.(2.30)) and
$R_{23}$ (Eq.(2.32)) respectively.
\par
In the limit $a \to 0$, the theory
reproduces the field equations for the Kerr-Newman
black hole\cite{chan}.  The solutions of the above equations
will determine the functions $A, B, \phi, \Psi, \omega$ and
$\rho$ as functions of $r$ and $\theta$ ($x_2$ and $x_3$).
\section{Mirror conjugate solutions}
Equations (2.40)-(2.46) possess alternative solutions which are
the mirror conjugates to those specified above.
This is a generalization of a symmetry of Ernst's equations for a
Kerr-Newman black hole\cite{chan}.
To see this we define the following transformation:
\begin{eqnarray}
& \tilde{t} \equiv \varphi \; ; \ \ \ \tilde{\varphi} \equiv -t&
\nonumber \\
&\tilde{\chi} \equiv -\strut\displaystyle{\chi\over{\chi^2 - \omega^2}}
\; ; \ \ \
\tilde{\omega} \equiv \strut\displaystyle{\omega\over{\chi^2 - \omega^2}}
& \nonumber \\
& -\tilde{A},_\mu \equiv \chi
\sqrt{\strut\displaystyle{\Delta\over{\delta}}}e^{-a\phi}
A,_r + \omega B,_\mu &\nonumber \\
& \tilde{A},_r \equiv \chi
\sqrt{\strut\displaystyle{\delta\over{\Delta}}}e^{-a\phi}
A,_\mu - \omega B,_r &\nonumber \\
& -\tilde{B},_\mu \equiv
\chi\sqrt{\strut\displaystyle{\Delta\over{\delta}}}e^{a\phi}
B,_r - \omega A,_\mu &\nonumber \\
& \tilde{B},_r \equiv \chi
\sqrt{\strut\displaystyle{\delta\over{\Delta}}}e^{a\phi}
B,_\mu + \omega A,_r &\nonumber \\
& \tilde{\phi} \equiv -\phi \; .
\end{eqnarray}
Also we define
\begin{eqnarray}
\tilde{\Psi} \equiv {\sqrt{\Delta\delta}\over{\tilde{\chi}}} \; .
\end{eqnarray}
The quantities $\rho, r, \theta, \Delta$ and $\delta$
are unchanged, and the field equations as well as the
metric are invariant under this transformation.
The metric is then
\begin{mathletters}
\begin{eqnarray}
ds^2 &=& -e^{2\tilde{\nu}}d\tilde{t}^2 +
e^{2\tilde{\psi}}(d\tilde{\varphi} - \tilde{\omega}d\tilde{t})^2 +
e^{2\mu_2}dr^2 + e^{2\mu_3}d\theta^2  \nonumber \\
& =& -e^{2\tilde{\nu}}d{\varphi}^2 +
e^{2\tilde{\psi}}(d{t} + \tilde{\omega}d{\varphi})^2 + e^{2\mu_2}dr^2 +
e^{2\mu_3}d\theta^2
\; .
\end{eqnarray}
and it can be re-written as follows,
\begin{eqnarray}
ds^2 &=& -\sqrt{\Delta \delta} \left[\chi dt^2 -
{1\over\chi} (d\varphi - \omega dt)^2\right]
+\rho^2\left[{dr^2\over\Delta} + d\theta^2\right]
\nonumber \\
& =&
-\sqrt{\Delta\delta}\left[\tilde\chi d\tilde t^2 -
{1\over\tilde\chi} (d\tilde\varphi - \tilde\omega d\tilde{t})^2\right]
+\rho^2\left[{dr^2\over\Delta} + d\theta^2\right]
\; ,
\end{eqnarray}
\end{mathletters}
which displays its invariance property.
\par
The invariance of the field equations under these transformations
means that for each solution $\Psi, \phi, \rho, \omega, A, B$ there
exists a dual solution $\tilde{\Psi}, \tilde{\phi}, \tilde{\rho}=\rho,
\tilde{\omega}, \tilde{A}, \tilde{B}$.
Since $\tilde{\phi} = -\phi$, the transformation in Eq.(3.1)
maps the weak gravitational coupling regime to the strong regime.
This is an example of mirror conjugation \cite{schw}.
The tilded quantities, following the convention of Chandrasekhar
\cite{chan}, are the ones calculated in the following section.
Eq.(3.1) can then be used to transform to the untilded quantities
if desired.
The inverse transformation, finally, is given as follows,
\begin{eqnarray}
&  \varphi \equiv \tilde{t} \; ; \ \ \ t \equiv -\tilde{\varphi}&
\nonumber \\
&\chi \equiv-\strut\displaystyle{\tilde\chi
\over{\tilde\chi^2 - \tilde\omega^2}}
\; ; \ \ \
\omega\equiv \strut\displaystyle{\tilde\omega
\over{\tilde\chi^2 - \tilde\omega^2}}&
\nonumber \\
& -A,_\mu \equiv \tilde\chi
\sqrt{\strut\displaystyle{\Delta\over{\delta}}}e^{-a\tilde\phi}
\tilde A,_r + \tilde\omega\tilde B,_\mu &\nonumber \\
& A,_r \equiv \tilde\chi
\sqrt{\strut\displaystyle{\delta\over{\Delta}}}e^{-a\tilde\phi}
\tilde A,_\mu - \tilde\omega\tilde B,_r &\nonumber \\
& -B,_\mu \equiv \tilde\chi
\sqrt{\strut\displaystyle{\Delta\over{\delta}}}e^{a\tilde\phi}
\tilde B,_r - \tilde\omega\tilde A,_\mu &\nonumber \\
& B,_r \equiv \tilde\chi
\sqrt{\strut\displaystyle{\delta\over{\Delta}}}e^{a\tilde\phi}
\tilde B,_\mu + \tilde\omega\tilde A,_r &\nonumber \\
& \phi \equiv -\tilde\phi \; .
\end{eqnarray}
\section{Perturbative expansion of the fields}
We now expand each of the fields in a power series of $Q^2$,
starting from the neutral Kerr solution.
This expansion will result in corrections to the fields in
terms of the square of the charge-to-mass ratio $(Q/M)^2$.
The fields we are expanding are the components of the
tilded fields of the convention used by Chandrasekhar \cite{chan}
as defined in Eq.(3.1a).
However for convenience we drop the tilde symbol in this section.
Also for convenience we introduce the following definitions
\begin{eqnarray}
\rho_0^2 &\equiv& r^2 + \alpha^2\mu^2 \nonumber \\
\Delta_0 &\equiv& r^2 - 2Mr + \alpha^2 \nonumber \\
\Psi_0 &\equiv& -{(\Delta_0 - \alpha^2\delta)
\over{\rho_0^2}}
\; ,
\end{eqnarray}
where the integration constants $M$ and $\alpha$
are identified from the leading order at large
$r$ with the mass and
the angular momentum per unit mass
of the Kerr black hole solution, and
\begin{eqnarray}
\Psi \equiv \Psi_0e^{-f} \; &;& \ \ \rho^2 \equiv
\rho_0^2e^{g}\; .
\end{eqnarray}
For later reference we note here that
\begin{eqnarray}
{\Psi,_{\sigma}\over{\Psi}} &=& {\Psi_0,_{\sigma}
\over{\Psi_0}} - f,_{\sigma}
\nonumber \\
{\rho,_{\sigma}\over{\rho}} &=& {\rho_0,_{\sigma}
\over{\rho_0}} + {1\over{2}} g,_{\sigma}
\; ,
\end{eqnarray}
where $\sigma=r$ or $\mu$.
\subsection{Series Expansion of the Field Equations}
The small charge perturbation expansions of the fields
are defined to be
\begin{eqnarray}
\omega,_{\sigma} &\equiv& e^f \sum_{n=0}^{\infty}
Q^{2n}\omega_{\sigma}^{(n)}
\nonumber \\
f,_{\sigma} &\equiv& \sum_{n=0}^{\infty} Q^{2n} f,_{\sigma}^{(n)}
\nonumber \\
g,_{\sigma} &\equiv& \sum_{n=0}^{\infty} Q^{2n} g,_{\sigma}^{(n)}
\nonumber \\
\phi,_{\sigma} &\equiv& \sum_{n=0}^{\infty} Q^{2n}
\phi,_{\sigma}^{(n)} \; ;
\ \ \ (\sigma=r\ {\rm or}\ \mu)
\ ,
\end{eqnarray}
and
\begin{eqnarray}
A,_{\sigma} &\equiv& e^{(a\phi-f)/2}\sum_{n=0}^{\infty}
Q^{2n-1}A_{\sigma}^{(n)}
\nonumber \\
B,_{\sigma} &\equiv& e^{-(a\phi+f)/2}\sum_{n=0}^{\infty}
Q^{2n-1}B_{\sigma}^{(n)}
\nonumber \\
\Delta &\equiv& \sum_{n=0}^{\infty} Q^{2n}\Delta^{(n)}
\; ;
\ \ \ (\sigma=r\ {\rm or}\ \mu)
\ .
\end{eqnarray}
When these expansions are inserted into Maxwell's equations,
the dilaton equation, the $\Psi$ equation, the $\omega$ equation and
the $\rho$ equations (Eqs.(2.40 - 2.46) respectively), the following
recursion relations are obtained for the n$^{th}$ order
expansion coefficients:
\par
\noindent
{\em Maxwell's equations:}
\begin{eqnarray}
\sum_{l+m=n}[(\Delta^{(m)} A_r^{(l)}),_r]
+ (\delta A_{\mu}^{(n)}),_{\mu} &=&
\sum_{l+m=n}\Delta^{(m)} A_r^{(l)}{\Psi_0,_r\over{\Psi_0}}
+ \delta A_{\mu}^{(n)} {\Psi_0,_{\mu}\over{\Psi_0}}
\nonumber \\
&&
+\Psi_0\sum_{l+m=n}\left[B_r^{(l)} \omega_{\mu}^{(m)}
- B_{\mu}^{(l)} \omega_r^{(m)} \right] \nonumber \\
&& + {1\over{2}} \sum_{l+m+k=n}
\Delta^{(k)} A_r^{(l)} (a\phi,_r^{(m)} - f,_r^{(m)})
\nonumber \\
&&+ {1\over{2}}\sum_{l+m=n}\delta A_{\mu}^{(l)}
(a\phi,_{\mu}^{(m)} - f,_{\mu}^{(m)})
\; ,
\end{eqnarray}
\begin{eqnarray}
\sum_{l+m=n}[(\Delta^{(m)} B_r^{(l)}),_r]
+ (\delta B_{\mu}^{(n)}),_{\mu} &=&
\sum_{l+m=n}[\Delta^{(m)} B_r^{(l)}{\Psi_0,_r\over{\Psi_0}}]
+ \delta B_{\mu}^{(n)} {\Psi_0,_{\mu}\over{\Psi_0}}
\nonumber \\
&& +\Psi_0\sum_{l+m=n}\left[-A_r^{(l)} \omega_{\mu}^{(m)}
+ A_{\mu}^{(l)} \omega_r^{(m)} \right] \nonumber \\
&& - {1\over{2}} \sum_{l+m+k=n}\left[\Delta^{(k)} B_r^{(l)} (a\phi,_r^{(m)}
+ f,_r^{(m)})\right]
\nonumber \\
&&-{1\over{2}}\sum_{l+m=n} \left[\delta B_{\mu}^{(l)}(a\phi,_{\mu}^{(m)}
+ f,_{\mu}^{(m)})\right]
\; ,
\end{eqnarray}
\par
\noindent
{\em Dilaton equation:}
\begin{eqnarray}
\sum_{l+m=n}&&\left[(\Delta^{(m)}\phi,_r^{(l)}),_r \right]
+ (\delta\phi,_{\mu}^{(n)})_{\mu}
= \nonumber \\
&&-{2a\over{\Psi_0}}\left[\sum_{l+m+k-1=n}\Delta^{(k)}(A_r^{(l)}
A_r^{(m)} - B_r^{(l)}B_r^{(m)}) + \sum_{l+m-1=n}
\delta(A_{\mu}^{(l)}
A_{\mu}^{(m)} - B_{\mu}^{(l)}B_{\mu}^{(m)})\right]
\; ,
\end{eqnarray}
\par
\noindent
{\em $\Psi$ equation:}
\begin{eqnarray}
\sum\limits_{l+m=n}\left[\Delta^{(l)}\delta
\left({\Delta^{(m)}\Psi_0,_r\over{\Psi_0}}\right),_r\right]
&+&\Delta^{(n)}\delta
\left({\delta\Psi_0,_{\mu}\over{\Psi_0}}\right),_{\mu}
- \sum_{l+m+k=n}\left[\Delta^{(k)}\delta
(\Delta^{(m)} f,_r^{(l)}),_r\right]\hfill \nonumber \\
-\sum\limits_{l+m=n}\left[\Delta^{(l)}\delta(\delta
f,_{\mu}^{(m)}),_{\mu}\right]
&=& -{2\over{\Psi_0}}\sum_{l+m+k+p-1=n} \Delta^{(p)}\delta
[\Delta^{(k)}(A_r^{(l)}
A_r^{(m)} + B_r^{(l)}B_r^{(m)})] \nonumber\\
&& -{2\over{\Psi_0}}\sum_{l+m+k-1=n} \Delta^{(k)}\delta
[\delta(A_{\mu}^{(l)}
A_{\mu}^{(m)} + B_{\mu}^{(l)}B_{\mu}^{(m)})]
\nonumber \\
&& {-\Psi_0^2} \sum_{l+m=n}\left[
\delta\omega_{\mu}^{(l)}\omega_{\mu}^{(m)}\right]
-\Psi_0^2 \sum_{l+m+k=n} \left[\Delta^{(k)}
\omega_r^{(l)}\omega_r^{(m)}\right]
\; ,
\end{eqnarray}
\par
\noindent
{\em $\omega$ equation:}
\begin{eqnarray}
\sum_{l+m=n}\left[\Delta^{(m)}\left(\Psi_0\omega_r^{(l)}
\right),_r\right] + \delta\left(\Psi_0\omega_{\mu}^{(n)}\right),_{\mu}
& =& {4\over{\Psi_0}}\sum_{l+m+k-1=n}\left[\Delta^{(k)}
\delta(A_{\mu}^{(l)}B_r^{(m)}
-A_r^{(l)}B_{\mu}^{(m)})\right]
\nonumber\\
&& - \sum_{l+m=n}\Delta^{(m)}\left[
\left({\Psi_0,_r\over{\Psi_0}}\right)
\left(\Psi_0\omega_r^{(l)}\right)\right]
\nonumber \\
&&- \left({\Psi_0,_{\mu}\over{\Psi_0}}\right)
\delta\left(\Psi_0\omega_{\mu}^{(n)}\right)
+\Psi_0\sum_{l+m=n}\left[\delta f,_{\mu}^{(l)}\omega_{\mu}^{(m)}\right]
\nonumber \\
&&+ \Psi_0\sum_{l+m+k=n}\left[\Delta^{(k)}f,_r^{(l)}
\omega_r^{(m)}\right]
\; ,
\end{eqnarray}
\par
\noindent
{\em $\rho$ equations:}
\begin{mathletters}
\begin{eqnarray}
\lefteqn{
\Delta^{(n)}\delta \left({\Psi_0,_r\over{\Psi_0}}\right)
\left({\Psi_0,_{\mu}\over{\Psi_0}}\right)
-\sum_{l+m=n}\Delta^{(l)}\delta\left[
\left({\Psi_0,_r\over{\Psi_0}}\right)f,_{\mu}^{(m)} +
\left({\Psi_0,_{\mu}\over{\Psi_0}}\right)f,_r^{(m)}\right]
+ {\Delta_r^{(n)}\over{2}}\delta,_{\mu}}
\nonumber \\
\lefteqn{+\sum_{l+m+k=n}\left[\Delta^{(k)}\delta f,_r^{(l)}
f,_{\mu}^{(m)}\right] -
\Psi_0^2\sum_{l+m=n}\omega_r^{(l)}\omega_{\mu}^{(m)}
-\delta\Delta,_r^{(n)}
\left({\rho_0,_\mu\over\rho_0}
+ {1\over 2}{\Psi_0,_{\mu}\over{\Psi_0}}\right)}
\nonumber \\
\lefteqn{- \delta,_{\mu}\Delta^{(n)}
\left({\rho_0,_r\over{\rho_0}} + {1\over{2}}
{\Psi_0,_r\over{\Psi_0}}\right)
-{1\over{2}}\sum_{l+m=n}\left[
\delta\Delta,_r^{(m)} (g,_{\mu}^{(l)}
- f,_{\mu}^{(l)}) + \delta,_{\mu}\Delta^{(m)}
(g,_r^{(l)} - f,_r^{(l)})\right]}
\nonumber      \\
&& = -{4\over{\Psi_0}}\sum_{l+m+k-1=n}\Delta^{(k)}\delta
(A_{\mu}^{(l)}A_r^{(m)} + B_r^{(l)}B_{\mu}^{(m)})
- \sum_{l+m+k=n}\delta\Delta^{(k)}\phi,_r^{(l)}\phi,_{\mu}^{(m)}
\end{eqnarray}
\begin{eqnarray}
\lefteqn{\Delta^{(n)}\delta\left[-2+{\delta\over{2}}\left({\Psi_0,_{\mu}
\over{\Psi_0}}\right)^2 -
\delta,_{\mu}\left({1\over{2}}{\rho_0,_{\mu}\over{\rho_0}}{\Psi_0,_{\mu}
\over{\Psi_0}}\right)\right]}
\nonumber    \\
\lefteqn{+ \sum_{l+m=n}\delta\Delta^{(m)}\left[-{\Delta^{(l)}\over{2}}
\left({\Psi_0,_r\over{\Psi_0}}\right)^2 + \Delta,_r^{(l)}
\left({\rho_0,_r\over{\rho}}+{1\over{2}}
{\Psi_0,_r\over{\Psi_0}}\right)\right]}
\nonumber    \\
\lefteqn{+ \sum_{l+m+k=n}\Delta^{(m)}\delta\Delta^{(k)}f,_r^{(l)}
\left({\Psi_0,_r\over{\Psi_0}}\right)
-\sum_{l+m=n}\Delta^{(m)}\delta^2
f,_{\mu}^{(l)}\left( {\Psi_0,_{\mu}\over{\Psi_0}}\right)}
\nonumber    \\
\lefteqn{+ {1\over{2}}\left[\sum_{l+m+k=n}\Delta^{(k)}\delta
\Delta,_r^{(m)}(g,_r^{(l)} - f,_r^{(l)}) -
\sum_{l+m=n}\Delta^{(m)}\delta\delta,_{\mu}(g,_{\mu}^{(l)} -
f,_{\mu}^{(l)})\right]}
\nonumber    \\
\lefteqn{+{1\over{2}}\left[-\sum_{l+m+k+p=n}\Delta^{(p)}\delta
\Delta^{(k)} (f,_r^{(l)}f,_r^{(m)})
+ \sum_{l+m+k=n}\Delta^{(k)}
\delta^2(f,_{\mu}^{(l)} f,_{\mu}^{(m)})\right]}
\nonumber    \\
\lefteqn{ + {\Psi_0^2\over{2}}\left[\sum_{l+m+k=n}\Delta^{(k)}
\omega_r^{(l)}\omega_r^{(m)}
-\sum_{l+m=n}\delta\omega_{\mu}^{(l)}\omega^{(m)}_{\mu} \right]}
\nonumber    \\
&& = {1\over{2}}\left[\sum_{l+m+p+k=n}\Delta^{(p)}\delta
\Delta^{(k)}\phi,_r^{(l)}
\phi,_r^{(m)} - \sum_{l+m+k=n}\Delta^{(k)}
\delta^2\phi,_{\mu}^{(l)}\phi,_{\mu}^{(m)}
\right] \nonumber \\
&& + {2\over{\Psi_0}}\left[
\sum_{l+m+k+p-1=n} \Delta^{(p)}\delta\Delta^{(k)}(A_r^{(l)}
A_r^{(m)}+B_r^{(l)}B_r^{(m)})
\right.\nonumber \\
&& \left.- \sum_{l+m+k-1=n}\Delta^{(k)}\delta^2(A_{\mu}^{(l)}
A_{\mu}^{(m)}+B_{\mu}^{(l)}B_{\mu}^{(m)})\right]
\end{eqnarray}
\begin{eqnarray}
\lefteqn{2\left[\Delta^{(n)}\delta \left(
{\delta\rho_0,_{\mu}\over{\rho_0}}\right),_{\mu}
+ \sum_{l+m=n} \Delta^{(m)}\delta
\left({\Delta^{(l)}\rho_0,_r\over{\rho_0}}\right),_r
\right]}
\nonumber    \\
\lefteqn{+ \left[\sum_{l+m=n}\Delta^{(m)}\delta(\delta
g,_{\mu}^{(l)}),_{\mu}+\sum_{l+m+k=n}\Delta^{(k)}
\delta(\Delta^{(m)}g,_r^{(l)}),_r\right]}
\nonumber    \\
&& = \Delta^{(n)}\delta\delta,_{\mu}\left(
{\rho_0,_{\mu}\over{\rho_0}} + {1\over{2}} {\Psi_0,_{\mu}
\over{\Psi_0}}\right) +
\sum_{l+m=n}\Delta^{(m)}\delta\Delta,_r^{(l)}\left({\rho_0,_r
\over{\rho_0}} +
{1\over{2}}{\Psi_0,_r\over{\Psi_0}}\right)
\nonumber \\
&& {1\over{2}}\left[\sum_{l+m+k=n}\Delta^{(k)}\delta
\Delta,_r^{(m)}(g,_r^{(l)} - f,_r^{(l)})
+ \sum_{l+m=n}\Delta^{(m)}\delta\delta,_{\mu}
(g,_{\mu}^{(l)} - f,_{\mu}^{(l)})\right]
\nonumber \\
&& - {1\over{2}}\left[
\sum_{l+m+k+p=n}\Delta^{(p)}\delta\Delta^{(k)}\phi,_r^{(l)}
\phi,_r^{(m)} +
\sum_{l+m+k=n}\Delta^{(k)}\delta^2
\phi,_{\mu}^{(l)}\phi,_{\mu}^{(m)}\right]
\nonumber \\
&& -{1\over{2}}\left[\sum_{l+m=n}\Delta^{(m)}\delta
\Delta^{(l)}\left({\Psi_0,_r
\over{\Psi_0}}\right)^2 + \Delta^{(n)}\delta^2
\left({\Psi_0,_{\mu}\over{\Psi_0}}\right)^2\right]
\nonumber \\
&&+ \sum_{l+m+k=n}\Delta^{(k)}\delta\Delta^{(m)} \left(
{\Psi_0,_r\over{\Psi_0}}\right)f,_r^{(l)} +
\sum_{l+m=n}\Delta^{(m)}\delta^2\left({\Psi,_{\mu}\over{\Psi}}\right)
f,_{\mu}^{(l)}
\nonumber \\
&& - {1\over{2}}\left[\sum_{l+m+k+p=n}\Delta^{(p)}\delta
\Delta^{(k)} f,_r^{(l)}
f,_r^{(m)} + \sum_{n=l+m+k}\Delta^{(k)}
\delta^2 f,_{\mu}^{(l)}f,_{\mu}^{(m)}\right]
\nonumber \\
&&+ {\Psi_0^2\over{2}} \left[\sum_{l+m=n}
\delta\omega_{\mu}^{(l)}\omega_{\mu}^{(m)} +
\sum_{l+m+k=n}\Delta^{(k)}
\omega_r^{(l)}\omega_r^{(m)}\right]
\end{eqnarray}
\end{mathletters}
\subsection{$n = 0$ Solutions}
Let us now find the solutions of these equations can be obtained
for the leading orders in the charge-to-mass ratio.
For $n=0$ we have the Kerr black hole characterized as follows
\begin{eqnarray}
A_{\sigma}^{(0)} &=& B_{\sigma}^{(0)} = 0  \\
\phi^{(0)} &=& 0
\; ,
\end{eqnarray}
as well as
\begin{eqnarray}
f^{(0)} = g^{(0)} = 0
\; ,
\end{eqnarray}
and
\begin{eqnarray}
\omega_r^{(0)} &=& -{2\alpha\delta M (r^2-\alpha^2\mu^2)
\over{(\Delta_0 - \alpha^2\delta)^2}}
\nonumber\\
\omega_{\mu}^{(0)} &=& - {2\alpha\Delta_0 M \mu r
\over{(\Delta_0 - \alpha^2\delta)^2}}
\nonumber\\
\Delta^{(0)}&=&\Delta_0 \; .
\end{eqnarray}
where $\Delta_0$ has been defined in Eq.(4.1).
\subsection{$n = 1$ Corrections}
At order $n=1$ the solutions of Maxwell's equations give
\begin{eqnarray}
&A^{(1)}_r =\strut\displaystyle{(-r^2+\alpha^2\mu^2)\over{\rho_0^4}}
\; ; \ \ \
A^{(1)}_{\mu} =-\strut\displaystyle{2\alpha^2\mu r \over{\rho_0^4}}&
\nonumber \\
& \alpha B^{(1)}_r = -A^{(1)}_{\mu} \; ; \ \ \
B^{(1)}_{\mu} = \alpha A^{(1)}_r
\; .&
\end{eqnarray}
From the dilaton equation we find
\begin{eqnarray}
\phi,_{\sigma}^{(1)}=- {a\over{M}} A^{(1)}_{\sigma} \; ;
\ \ \ (\sigma=r\ {\rm or}\ \mu)
\ .
\end{eqnarray}
To obtain the remaining quantities we begin with the
assumption that $\Delta^{(1)} = \Delta^{(1)}(a) =$ constant.
At first we assume the following ansatz
\begin{eqnarray}
f^{(1)} = \Delta^{(1)} \hat f^{(1)} \; ; \ \ \
g^{(1)} = \Delta^{(1)} \hat g^{(1)} \; ; \ \ \
\omega^{(1)}_{\sigma} = \Delta^{(1)}\hat \omega^{(1)}_\sigma
\; .
\end{eqnarray}
Now when we set the dilaton parameter $a=0$ in the field equations
(2.2)-(2.5), we recover the Kerr--Newman space--time for which an
exact solution is known \cite{chan}.
For such a case, we therefore have $\Delta^{(1)}(a=0) = 1$,
$\Delta^{(n)} = 0$ $(n>1)$ and the quantities
$\hat f^{(1)}, \hat g^{(1)}, \hat \omega^{(1)}_\sigma,
\omega^{(0)}_\sigma, A^{(1)}_{\sigma} $ and $B^{(1)}_{\sigma}$
are order $n=1$ (i.e. $Q^2$) solutions of the Kerr--Newman field
equations.
The exact solutions for the Kerr--Newman case are
\begin{eqnarray}
&\hat \Psi = -\strut\displaystyle{\left[(\Delta_0 + Q^2) -
\alpha^2\delta\right]\over{\rho_0^2}} \equiv \Psi_0 e^{-\hat f}
\ ; \ \ \
\hat f=-\ln\left(1-\strut\displaystyle{Q^2\over\rho_0^2 \Psi_0}\right)
\ ,&
\nonumber \\
&\hat \rho^2 = r^2 + \alpha^2\mu^2 \equiv \rho_0^2 e^{\hat g}
\ ;\ \ \
\hat g=0&  \\
&\hat A,_\sigma= Q A_\sigma^{(1)}
\ ;\ \ \
\hat B,_\sigma= Q B_\sigma^{(1)}
\ ,&   \\
&\hat \omega = -\alpha\delta[1+\hat \Psi^{-1}]&
\nonumber \\
&\hat \Psi\hat \omega,_r = \alpha\delta\left[
\strut\displaystyle{\Psi_0,_r\over{\Psi_0}} - \hat f,_r\right]&
\nonumber \\
&\hat \Psi\hat \omega,_{\mu} = 2\alpha\mu[1 + \hat \Psi] +
\alpha\delta\left[\strut\displaystyle{\Psi_0,_{\mu}
\over{\Psi_0}} - \hat f,_\mu\right]
\ .&
\end{eqnarray}
The functions $\hat f, \hat g$ and $\hat \omega,_{\sigma}$ can
be expanded as was done for their counterparts in Eq.(4.4).
From Eqs.(4.9) with $a=0$, we must have
\begin{eqnarray}
& \hat f^{(1)} = \strut\displaystyle{1\over{\rho_0^2\Psi_0}}&
\nonumber \\
& \hat g^{(1)} = 0 &
\end{eqnarray}
and
\begin{eqnarray}
&\Psi_0\hat \omega^{(1)}_{r} = -\alpha\delta \hat f,^{(1)}_r&
\nonumber \\
&\Psi_0\hat \omega^{(1)}_{\mu} = -2\alpha\mu\Psi_0 \hat f^{(1)}
- \alpha\delta \hat f,^{(1)}_\mu&
\end{eqnarray}
On the other hand, if we set $\Delta^{(1)} =0$, which corresponds to
the Kaluza--Klein case $a^2=3$ \cite{horne,gibb},
we find the solutions
\begin{eqnarray}
&\bar{f}^{(1)} = \bar{g}^{(1)} =
\strut\displaystyle{A^{(1)}\over{M}}&
\nonumber \\
&\Psi_0\bar{\omega}^{(1)}_r = -\alpha\delta
\bar{f},^{(1)}_r& \nonumber \\
&\Psi_0\bar{\omega}^{(1)}_\mu =
-\strut\displaystyle{\Delta_0 \bar{f},^{(1)}_\mu\over{\alpha}}
\ ,&
\end{eqnarray}
where $A^{(1)} \equiv r/\rho^2_0$.
Thus for general $\Delta^{(1)} = \Delta^{(1)}(a)$,
we attempt solutions of the form
\begin{eqnarray}
&f^{(1)} = \strut\displaystyle{\Delta^{(1)}\over{\rho_0^2\Psi_0}}
+ \lambda_0\strut\displaystyle{A^{(1)}\over{M}}
\ ; \ \ \
g^{(1)}= \lambda_0\strut\displaystyle{A^{(1)}\over{M}} &
\nonumber \\
&\Psi_0\omega^{(1)}_r = -\alpha\delta f,^{(1)}_r&
\nonumber \\
&\Psi_0\omega^{(1)}_{\mu} = -2\alpha\mu\Psi_0 f^{(1)}
- \alpha\delta f,^{(1)}_\mu
\ ,
&
\end{eqnarray}
where $\lambda_0$ is a function of $a$ to be determined,
as is $\Delta^{(1)}$.
Substituting these expressions into the recursion formulas
of subsection A. for $n=1$, and modding out the Kerr-Newman part, we find
\begin{eqnarray}
\Delta^{(1)}(a) = 1 - \lambda_0 \ .
\end{eqnarray}
Note that this form reproduces the two previous limiting cases
$a=0$ and $a=\sqrt{3}$ when $\lambda_0=a^2/3$.
The proof of this conjecture will be presented at the order $n=2$
calculation in the next subsection.
\par
In summary, the contributions to the fields at this order are given
as
\begin{eqnarray}
&A_r^{(1)} = \strut\displaystyle{(-r^2 + \alpha^2\mu^2)\over{\rho_0^4}}
\; ; \ \ \
 A_\mu^{(1)} =-\strut\displaystyle{2 \alpha^2\mu r\over{\rho_0^4}}&
\nonumber \\
&\alpha B_r^{(1)} = - A^{(1)}_\mu \; ; \ \ \ B_\mu^{(1)} =
\alpha A^{(1)}_r&
\nonumber \\
&\phi,_\sigma^{(1)} = -\strut\displaystyle{a A^{(1)}_\sigma\over{M}}
\ ; \ \ \ (\sigma=r\ {\rm or}\ \mu)
& \nonumber \\
& f^{(1)} = \strut\displaystyle{\Delta^{(1)}\over{\rho_0^2\Psi_0}}
+ \lambda_0\strut\displaystyle{A^{(1)}\over{M}} \ ; \ \ \
g^{(1)}= \lambda_0\strut\displaystyle{A^{(1)}\over{M}} &
\nonumber \\
&\Psi_0\omega^{(1)}_{r} = -\alpha\delta f,^{(1)}_r \ ; \ \ \
\Psi_0\omega^{(1)}_{\mu} = -2\alpha\mu\Psi_0 f^{(1)}
- \alpha\delta f,^{(1)}_\mu
\ ,
&
\end{eqnarray}
with $\Delta^{(1)}$ given by Eq.(4.26).
\subsection{$n = 2$ Corrections}
We have also obtained the order $n=2$ (i.e. $Q^4$)
corrections to the electromagnetic fields.
To obtain these corrections we begin by noting that the
exact solutions for the Kerr--Newman case are:
\begin{eqnarray}
&\hat A,_r = Q\strut\displaystyle{(-r^2+\alpha^2\mu^2)\over{\rho_0^4}}
\; = \; e^{-\hat f/2} \sum\limits_{n=0}^\infty Q^{2n-1} \hat A_r^{(n)}
& \nonumber \\
&\hat A,_\mu = -Q\strut\displaystyle{2\alpha^2\mu r\over{\rho_0^4}}
\; = \; e^{-\hat f/2} \sum_{n=0}^\infty Q^{2n-1} \hat A_\mu^{(n)}
& \nonumber \\
& \hat B,_\sigma = e^{-\hat f/2} \sum\limits_{n=0}^\infty Q^{2n-1}
\hat B_{\sigma}^{(n)}
\ ; \ \ \ (\sigma=r\ {\rm or}\ \mu)
\; . &
\end{eqnarray}
From Eqs.(4.16) and (4.20) we have
\begin{eqnarray}
\hat A,_\sigma = Q A_\sigma^{(1)} \; ; \ \ \
\hat B,_\sigma = Q B_{\sigma}^{(1)}
\; ,
\end{eqnarray}
which imply that
\begin{eqnarray}
&\hat A_{\sigma}^{(2)} = {1\over{2}} \hat f^{(1)} A_{\sigma}^{(1)}&
\nonumber \\
&\hat B_{\sigma}^{(2)} = {1\over{2}} \hat f^{(1)} B_{\sigma}^{(1)}&
\; ,
\end{eqnarray}
where again $\hat f^{(1)} = 1/\rho_0^2\Psi_0$ (Eq.(4.22)).
This suggests that we try the following ansatz
\begin{eqnarray}
&A_\sigma^{(2)} = \Delta^{(1)} \hat A_{\sigma}^{(2)} +
\lambda_0 c A_\sigma^{(1)}&
\nonumber \\
&B_\sigma^{(2)} = \Delta^{(1)} \hat B_{\sigma}^{(2)} +
\lambda_0 h B_\sigma^{(1)}&
\ .
\end{eqnarray}
Substituting these expressions into Maxwell's equations (Eqs.(4.6,4.7))
for $n=2$ and modding out the Kerr--Newman part, the equations are satisfied
when $c$ and $h$ are given as
\begin{eqnarray}
c &=&- {\Psi_0\over{4M^2}}\left(1+{a^2\over{\lambda_0}}\right)
\nonumber \\
h &=& {\Psi_0\over{4M^2}}\left(1-{a^2\over{\lambda_0}}\right)
\; ,
\end{eqnarray}
provided that
\begin{eqnarray}
\lambda_0 = {a^2\over{3}}
\; ,
\end{eqnarray}
which is the expected form.
The order $n=2$ corrections for the $A$ and $B$ potentials are
therefore
\begin{eqnarray}
&A_\sigma^{(2)} =
\left(\strut\displaystyle{\Delta^{(1)}\over{2\rho_0^2\Psi_0}}
- \strut\displaystyle{a^2\Psi_0 \over{3M^2}}\right) A_\sigma^{(1)}
& \nonumber \\
&B_\sigma^{(2)} =
\left(\strut\displaystyle{\Delta^{(1)}\over{2\rho_0^2\Psi_0}}
- \strut\displaystyle{a^2\Psi_0 \over{6M^2}}\right) B_\sigma^{(1)}
\ ,
&
\end{eqnarray}
with $\Delta^{(1)} = 1 - a^2/3$.
\subsection{Fields and Metric Tensor Elements}
In this section we summarize the expressions for the electric,
magnetic and dilaton fields and the expressions for the
tensor elements.
Using the definitions
\begin{eqnarray}
C_A &\equiv& {\Delta^{(1)}\over{2\,\rho^2_0\,\Psi_0}} -
{a^2\,\Psi_0 \over{3\,M^2}}
\nonumber\\
C_B &\equiv& {\Delta^{(1)}\over{2\,\rho^2_0\,\Psi_0}} -
{a^2\,\Psi_0 \over{6\,M^2}}
\end{eqnarray}
we have for the electric and magnetic field in tetrad components
with respect to untilded coordinates (recalling that the quantities
we previously calculated are the tilde ones),
\begin{eqnarray}
{\cal E}_{\hat{\varphi}} &=& 0
\nonumber \\
{\cal E}_{\hat{r}} &=& F_{\hat{t}\hat{r}} = -\tilde{ A},_r
\nonumber \\
&=& -Q\, \tilde{A}^{(1)}_r\,[1 + C_A\, Q^2]
\nonumber \\
{\cal E}_{\hat{\theta}} &=& F_{\hat{t}\hat{\theta}}
= {\sqrt{\delta}\, \tilde{A},_\mu\over{r}}
\nonumber \\
&=& {\sqrt{\delta}\, \tilde{A}^{(1)}_\mu\over{r}}
\, [1 + C_A Q^2]
\end{eqnarray}
and
\begin{eqnarray}
{\cal B}_{\hat{\varphi}} &=& 0
\nonumber\\
{\cal B}_{\hat{r}} &=& F_{\hat{\varphi}\hat{\theta}}
= {\sqrt{\delta}\, B,_\mu\over{r^2\sin\theta}}
\nonumber\\
&\approx & -{\sqrt{\delta}\Delta\over{\tilde\Psi}}\,
{e^{a\tilde{\phi}}\, Q
\tilde{B}^{(1)}_r\over{r^2\sin\theta}}\,[1 + C_B\, Q^2]
\nonumber \\
{\cal B}_{\hat\theta} &=& -F_{\hat{\varphi}\hat{r}}
 = {B,_r\over{r\sin\theta}}
\nonumber\\
&\approx &
{\delta\over{\tilde\Psi}}\,
{e^{a\tilde{\phi}}\tilde{B}^{(1)}_\mu\over{r\sin\theta}}
\,[1 + C_B\, Q^2] \; .
\end{eqnarray}
In the limit $r \to \infty$ the fields can be approximated as
\begin{eqnarray}
{\cal E}_{\hat{r}} &\approx& {Q_{phys}\over{r^2}}
\nonumber \\
{\cal E}_{\hat{\theta}} &\approx& -{2\,\alpha^2\over{r^4}}\, Q_{phys}\,
\sin\theta\,\cos\theta
\nonumber \\
{\cal B}_{\hat{r}} &\approx& {2\,\alpha\, Q_{phys}\over{r^3}}\,\cos\theta
\,\left[1 - {a^2\,Q^2\over{6\,M^2}}\right]
\nonumber \\
{\cal B}_{\hat{\theta}} &\approx& {\alpha\, Q_{phys}\over{r^3}}\,\sin\theta
\left[1 - {a^2\,Q^2\over{6\,M^2}}\right]
\; .
\end{eqnarray}
where
\begin{eqnarray}
Q_{phys} \equiv Q\,\left[1 + {a^2\,Q^2\over{3\,M^2}}\right]
\ ,
\end{eqnarray}
and the hatted variables are the usual flat spacetime
spherical coordinates\cite{misner}.
If we square the latter relation, we find that to
order $Q^3, \ \ Q^2_{phys} = Q^2$.
\par
To order $Q^2$ the dilaton field solution agrees with the following
expression
\begin{eqnarray}
\tilde\phi = -{1\over{a}}\,\ln{b^{2\,\gamma}}
\end{eqnarray}
where
\begin{eqnarray}
\gamma = {2\,a^2\over 1 + a^2}
\; ,
\end{eqnarray}
and
\begin{eqnarray}
b^2 = 1 + {Q^2\,(1+a^2)\,M\,r \over 2\,M^2\,\rho_0^2}
\; .
\end{eqnarray}
When $a^2=3$ this form reproduces the Kaluza--Klein solution
\cite{horne,gibb}.
\par
Also, our perturbative result for $\tilde\Psi$ and $\rho$ agrees
with the following expressions (up to order $Q^2$)
\begin{eqnarray}
\tilde\Psi=-{\Delta-\alpha^2\,\delta\over\rho^2}
\ ,
\end{eqnarray}
and
\begin{eqnarray}
\rho^2=\rho_0^2\,e^{-a\,\tilde\phi/3}
\ .
\end{eqnarray}
\par
Similarly, the following expression
\begin{eqnarray}
\tilde\omega=-\alpha\,\delta\,[1+\tilde\Psi^{-1}]
\ ,
\end{eqnarray}
reproduces our perturbative results to order $Q^2$.
\par
Let us notice further that the following expressions for the potentials
$\tilde A$ and $\tilde B$ are again in agreement with our
perturbative calculation to order $Q^2$
\begin{eqnarray}
\tilde A=Q_{phys}\,{r\over\rho^2}\,e^{a\,\tilde\phi/3}
\ ;\ \ \
\tilde B=-Q_{phys}\,{\alpha\,\mu\over\rho^2}\,e^{-a\,\tilde\phi/3}
\,\left(1-{a^2\,Q^2\over6\,M^2}\right)
\ .
\end{eqnarray}
\par
The covariant and contravariant metric tensor elements
to the same order in $Q$ are (we have reinstated the tilde)
\begin{eqnarray}
g_{tt} &=& \tilde \Psi
\nonumber\\
&\approx& -\left[1 - {2M\over{r}}\left({1 + {Q^2\,
\lambda_0\over 2M^2}}\right)\right]
\nonumber \\
g^{tt} &=& {1\over\tilde \Psi} - {\tilde \omega^2\, \tilde{\Psi}
\over{\Delta\delta}}
\nonumber\\
&\approx& -\left[1 + {2M\over{r}}
\left(1 + {Q^2\,\lambda_0 \over{2M^2}}\right)\right]
\nonumber\\
g_{\varphi\varphi} &=& \tilde \Psi\,\tilde \omega^2-
{\Delta\delta \over{\tilde \Psi}}
\nonumber \\
&\approx& r^2\, \delta \nonumber \\
g^{\varphi\varphi} &=& -{\tilde \Psi\over{\Delta\,\delta}}
\nonumber\\
&\approx& {1\over{r^2\delta}} \nonumber \\
g_{t\varphi} &=& \tilde \Psi\,\tilde \omega
\nonumber \\
&\approx& {-2  \alpha\,\delta\, M\over{r}}
\left(1 + {Q^2\,\lambda_0\over{2\,M^2}}\right)
\nonumber \\
g^{t\varphi} &=& {1\over{\Delta\,\delta}}\,\tilde\Psi\,\tilde \omega
\nonumber \\
&\approx& {-2\alpha\,M \over{r^3}}\left(1 +
{Q^2\,\lambda_0\over{2\,M^2}}\right) \nonumber \\
g_{rr} &=& {\rho^2 \over{\Delta}}
\nonumber \\
&\approx& 1 + {2M\over{r}}\left( 1 + Q^2\,
{\lambda_0\over{2\,M^2}} \right) \nonumber \\
g^{rr} &=& {\Delta\over{\rho^2}}
\nonumber \\
&\approx& 1 - {2\,M\over{r}}
\left(1 + {Q^2\lambda_0\over{2\,M^2}}\right) \nonumber \\
g_{\theta\theta} &=& \rho^2
\nonumber \\
&\approx& r^2 \nonumber \\
g^{\theta\theta} &=& \rho^{-2}
\nonumber \\
&\approx& r^{-2} \; ,
\end{eqnarray}
where $\approx$ denotes again the $r\to\infty$ approximation.
\section{Corrections to the Gyromagnetic Ratio}
The gyromagnetic ratio for the Kerr-Newman black hole is known
to be $g = 2$ \cite{strau}.
The dilaton component of the gravitational field shifts $g$ away
from 2.
The shift has been calculated for the case of small black hole
angular momentum\cite{horne}.
Using the expressions for the $g_{t\varphi}$ component of the
tensor and the ${\cal B}_\theta$ component of the magnetic field,
we can calculate the gyromagnetic ratio for arbitrary angular
momentum.
The magnetic moment measured by a distant observer is
\begin{eqnarray}
\mu_{phys} = g\,{Q_{phys}\, J_{phys}\over{2\,M_{phys}}}
\ ,
\end{eqnarray}
where $Q_{phys}$ is given in Eq.(4.39) and  $\mu_{phys}$,
$J_{phys}$ and $M_{phys}$ are obtained from the asymptotic
expressions for ${\cal B}_\theta$, $g_{t\varphi}$ and $g_{tt}$
respectively
\begin{eqnarray}
\mu_{phys} &=& \alpha\, Q\, \left[ 1 + {a^2\,Q^2\over{6\,M^2}}\right]
\nonumber \\
J_{phys} &=& \alpha\, M\, \left[ 1 + {a^2\,Q^2\over{6\,M^2}}\right]
\nonumber \\
M_{phys} &=& M\, \left[ 1 + {a^2\,Q^2\over{6\,M^2}}\right]
\end{eqnarray}
and, as we stated previously, all corrections are proportional to
the $(Q/M)^2$ ratio.
The gyromagnetic ratio obtained from these expressions is
\begin{eqnarray}
g = 2\,\left[ 1 - {a^2\,Q_{phys}^2\over{6\,M^2_{phys}}}\right]
\ .
\end{eqnarray}
This agrees with the expression obtained in Ref.\cite{horne}
when their expression is expanded to order $Q^2$.
Our expression differs from theirs in that the ansatz they chose
for the metric assumes there is no dilatonic correction to the
physical mass and the charge.
Our solutions contain no assumptions, other than requiring that
our solutions give the Kerr--Newman solutions in the $a=0$ limit,
and predict that both the charge and the mass of the black hole
are modified by the presence of the dilaton.
\section{Discussion}
The solutions obtained here for the fields should be very useful
in determining the effects of a scalar gravitational field on
physically measurable quantities.
In our derivation we have made no assumptions about the form of
the metric tensor elements other than the small charge-to-mass ratio of
the black hole.
In particular our solutions are valid for black holes with arbitrary
angular momentum and dilatonic charge $a$,
whereas previous solutions have required that the black
holes have small angular momentum or specific values of $a$.
These are important distinctions, because a black hole
can in principle have appreciable angular momentum.
Having $a$ as an arbitrary parameter is, of course, a desirable
feature of any measurable quantity obtained for this geometry.
\par
Our solutions provide a framework for the calculation of
physically measurable quantities, such as radiation intensities
for radiation emitted from the region near a black hole horizon.
Since the wave equation and Maxwell's
equations are separable in the Kerr--Newman geometry \cite{misner},
it may be possible to obtain analytic expressions for the
wave functions, at least in some linear approximation of the
gravitational field.
We are presently carrying out these calculations.
The goal of these calculations is to provide measurable
predictions of the effect of the scalar component of gravity.
\par
Having explicit expressions for the metric tensor elements will
allow us to carry out an analysis of the statistical
mechanics of a gas of Kerr--Newman dilatonic black holes as we have
done previously for the Schwarzschild, Reissner-Nordstr\o m
and dilaton black holes \cite{hl1}.
In particular it will be interesting to see if a gas of such
black holes obeys the bootstrap relation.
\par
Finally, we note here that our solutions possess a mirror symmetry.
This symmetry has been studied extensively within
the context of string theory, where it has been conjectured to be
a property of the Calabi-Yau space associated with a
(2,2) conformal field theory.
It may then be an interesting problem to generalize the present
work to the case where both a dilaton and an axion are present.
\acknowledgments
This work was supported in part by the U.S. Department of Energy
under Grant No. DE-FG05-84ER40141

\end{document}